\documentclass[11pt,a4paper,aip,jmp,nofootinbib]{revtex4-1}
\usepackage{amsmath,amssymb,enumerate,epsfig,varioref}
\usepackage{hyperref}

\newtheorem{proof}{Proof}

\long\def\nodo#1{}
\def\id{\mathrm{id}}
\def\DD{\mathcal{D}}

\def\CC{\mathcal{C}}
\def\C{\mathbb{C}}

\def\F{\mathcal{F}}
\def\g{\mathfrak{g}}
\def\gg{\mathfrak{g}}
\def\ggx{\mathfrak{g}_{\hat{x}}}
\def\ggy{\mathfrak{g}_{\hat{y}}}

\def\p{\partial}

\def\al{\alpha}

\def\PL#1{Phys.\ Lett.\ {\bf#1}}

\def\PR#1{Phys.\ Rev.\ {\bf#1}}

\def\JoP#1{J.\ Phys.\ {\bf#1}} \def\IJMP#1{Int.\ J. Mod.\ Phys.\ {\bf #1}}

\def\EPJ#1{Eur.\ Phys.\ J.\ {\bf#1}}
\newcommand{\arxiv}[1]{\href{https://arxiv.org/abs/#1}{arXiv:#1}}
\newcommand{\bibx}[3]{#1, #2, \arxiv{#3}}
\newcommand{\bibxp}[4]{#1, #2, #3, \arxiv{#4}}
\newcommand{\bibp}[3]{#1, #2, #3}

\begin{document}

\title{Symmetric ordering and Weyl realizations for quantum Minkowski spaces}

\author{Stjepan Meljanac}
\affiliation{Rudjer Bo\v{s}kovi\'{c} Institute, Theoretical
Physics Division, Bijeni\v{c}ka c. 54, HR 10002 Zagreb, Croatia}
\email{meljanac@irb.hr}

\author{Zoran \v{S}koda}
\affiliation{Department of Teacher's Education, University of
Zadar, F. Tudjmana 24, 23000 Zadar, Croatia}
\email{zskoda@unizd.hr}

\author{Sa\v{s}a Kre\v{s}i\'{c}--Juri\'{c}}
\affiliation{Faculty of Science, University of Split, Rudjera
Bo\v{s}kovi\'{c}a 33, 21000 Split, Croatia}
\email{skresic@pmfst.hr}


\begin{abstract}
Symmetric ordering and Weyl realizations for non commutative quantum
Minkowski spaces are reviewed. Weyl realizations of Lie deformed
spaces and corresponding star products, as well as twist corresponding
to Weyl realization and coproduct of momenta are presented.
Drinfeld twists understood in Hopf algebroid sense are also discussed.
A few examples of corresponding Weyl realizations are given.
We show that for the original Snyder space there exists symmetric
ordering, but no Weyl realization. Quadratic deformations of Minkowski
space are considered and it is demonstrated that symmetric ordering
is deformed and a generalized Weyl realization can be defined.
\end{abstract}


\maketitle


\section{Introduction}

Snyder model~\cite{Snyder} was the first example of a noncommutative (NC) geometry based on a deformation of the Heisenberg algebra.
Noncommutative spaces were introduced to better understand and model Planck scale phenomena and to obtain quantum gravity models
that reconcile general relativity and quantum mechanics~\cite{dfr,garay} as well as to model deformations of quantum mechanical phase space
algebra~\cite{yang}. A widely studied model is the $\kappa$-Minkowski spacetime, where the parameter $\kappa$ is usually interpreted as
the Planck mass or the quantum gravity scale and the coordinates close a Lie algebra~\cite{luknowruto,luknowru,majidru,gacluknow,dasklukwor}.

A successful approach to symmetries in noncommutative geometry is based on the formalism of Hopf algebras~\cite{majid,chapre}.
In the Hopf algebra approach, it is possible to deform the Hopf algebra using a twist element satisfying the 2-cocycle condition.
It leads to a new Hopf algebra, with unchanged algebra sector and a deformed coalgebra sector. This approach describes relativistic
symmetries of the quantum spacetime~\cite{arzkowglik}.  The $\kappa$-Poincar\'e quantum group~\cite{maslanka,zakrzewski,gillkosmajmasku}
is a possible quantum symmetry of the $\kappa$-Minkowski spacetime allowing to study deformed relativistic spacetime symmetries and the
corresponding dispersion relations~\cite{acmajid,gac,gac18}. Deformations of relativistic symmetries play an important role in the study of
phenomenologically relevant effects of quantum gravity~\cite{gac18,gac19,magsmol,kowgliknow,gacsmolstar,ballgubmer}.

A powerful tool in the study of NC spaces is that of realizations in terms of the Heisenberg
algebra~\cite{beckers,dong,meljkjst,meljsamstgup,govgupharmelj,govgupharmelj2,battisti1,battisti2,meljk-j,meljsamst2}
where NC coordinates are expressed in terms of commutative coordinates and the corresponding momenta. This allows one to simplify
the methods of calculation on the deformed spacetime. Every realization corresponds to a specific ordering, an important example
being the Weyl realization related to the symmetric ordering. Exponential formulas~\cite{battisti2,meljsamst2,meljjord2017,mstrajnexp21}
are related to the deformed coproduct of momenta in NC spaces as well as to computations of star products. They are important for
construction of a field theory, the notion of a differential calculus and other calculations in a NC spacetime~\cite{diffc,madore,gacarz}.
Lie deformed phase spaces obtained from twists in the Hopf algebroid sense were considered in Refs. \onlinecite{meljjord2017,mstrajnexp21,jurmeljs,jurkovmelj,lukmeljpikwor}.

A class of deformed quantum phase spaces, i.e.\ a deformed Heisenberg algebra, is generated with NC coordinates $\hat x_\mu$ and commutative momenta $p_\mu$ defined as 
(see for example Ref. \onlinecite{ms2112})
\begin{align}\label{eq:thetagen}
[\hat{x}_\mu, \hat{x}_\nu] &= i\theta_{\mu\nu}(l,\hat{x}, p),&& \\
\left[p_\mu,\hat{x}_\nu\right] &=-i\varphi_{\mu\nu}(l,p),&& \\
\left[p_\mu, p_\nu\right] &=0,
\end{align}
where all Jacobi identities are satisfied and $l$ is a real parameter on the order of the Planck length. The undeformed phase space is defined as
\begin{equation}
[x_\mu, x_\nu] = 0,\\
\left[p_\mu, p_\nu\right] = 0,\\
\left[x_\mu, p_\nu\right] = -i\eta_{\mu\nu},\,\,\,\,\mu,\nu = 0, ..., n-1,
\end{equation}
where $\eta_{\mu\nu} = \text{diag}(-1, 1, ..., 1)$, $x_\mu$ are coordinates and $p_\mu$
are momenta. There is a Fock action $\triangleright$ of undeformed phase space algebra on polynomials, and more generally on formal functions of $x_\mu$, determined by the formulas
\begin{eqnarray}\label{eq:Fock}
x_\mu\triangleright f(x) = x_\mu f(x),\\
p_\mu\triangleright f(x) = -i \frac{\partial f(x)}{\partial x_\mu}.
\end{eqnarray}

Examples of NC spaces where $\theta_{\mu\nu}$ in~(\ref{eq:thetagen}) do not depend on momenta $p_\mu$ are canonical theta space with $\theta_{\mu\nu} = \mathrm{const}$~\cite{dfr},
Lie algebra type spaces with $\theta_{\mu\nu} = C_{\mu\nu\lambda}\hat{x}_\lambda$~\cite{luknowruto,luknowru,majidru,gacluknow,dasklukwor,gacarz,lukwor,daskwal,lukmeljpikwor}
and quadratic deformations of Minkowski space with  $\theta_{\mu\nu} =  \theta_{\mu\nu\rho\lambda}\hat{x}_\rho \hat{x}_\lambda$~\cite{Wess,lukwor}.
In the Snyder space~\cite{Snyder,mig,migc,girliv,lustern1,lustern2} where $[\hat x_\mu, \hat x_\nu] = i l^2 M_{\mu\nu}$ the NC coordinates $\hat{x}_\mu$
do not close an algebra between themselves, but $\hat{x}_\mu$ and the Lorentz generators $M_{\mu\nu}$ close a Lie algebra. Some new developments in
applications of NC geometry to physics can be found in Refs. \onlinecite{kupvit,kupkurvit,lizzvit,matwall1,matwall2}.

The symmetric ordering of the product
$\hat{x}_{\mu_1}\cdots\hat{x}_{\mu_N}$ of $N$ elements
(possibly with repetitions) of
a basis $\hat{x}_0,\ldots,\hat{x}_{n-1}$ of a Lie algebra $\ggx$
of dimension $n$ is an element in its enveloping algebra $U(\ggx)$ defined as
the sum over all permutations
\begin{equation}\label{eq:sym}
\frac{1}{N!}\sum_{\sigma\in \Sigma_N}\hat{x}_{\mu_{\sigma(1)}}\cdots\hat{x}_{\mu_{\sigma(N)}},
\end{equation}
where $\Sigma_N$ denotes the permutation group on $N$ letters. Denote by
$S(\gg)$ the symmetric algebra of the underlying vector space $\gg$ of the Lie algebra $\ggx$.
The symmetric algebra may be viewed as
the algebra of polynomials in coordinates
$x_0,x_1,\ldots,x_{n-1}$ which correspond to the same basis $\hat{x}_0,\hat{x}_1,\ldots,\hat{x}_{n-1}$, but mutually commute.
The symmetrization map (see Refs. \onlinecite{bourbaki}, Ch.~II; \onlinecite{petracci}; \onlinecite{Durov}, Ch. 10) is the isomorphism
of vector spaces ({\em a fortiori}, of coalgebras) mapping
a monomial $x_{\mu_1}\cdots x_{\mu_N}$ in the symmetric algebra $S(\gg)$
to the symmetrically ordered expression~(\ref{eq:sym}) in $U(\ggx)$.
The Weyl realization of $U(\ggx)$ is the homomorphic
realization of $U(\ggx)$ within the undeformed Heisenberg algebra
(expressing the generators $\hat{x}_\mu$
in terms of coordinates $x_\nu$ and momenta $p_\nu$), which satisfies
\begin{equation}\label{eq:weyl}
  (k_\alpha\hat{x}_\alpha)^N\triangleright 1 = (k_\alpha x_\alpha)^N
\end{equation}
for all $k = (k_0,k_1,\ldots,k_{n-1})$ and all $N\geq 1$.
The assignment $\hat{f}\mapsto \hat{f}\triangleright 1$,
where the right-hand side is computed in the Weyl realization,
defines a map $U(\ggx)\to S(\gg)$
which is precisely the inverse of the symmetrization map.

Weyl realizations\footnote{Though the term {\it Weyl realization} came from a vague analogy with Weyl ordering and Weyl quantization of
observables which introduces symmetrized combinations of coordinates with corresponding momenta operators when building quantized observables,
both physics and mathematics are quite different here. Physically, our symmetrization is among coordinates only. Mathematically, related formulas and context are rather different.}
related to Lie algebras were studied in Refs. \onlinecite{meljkjst,petracci,Durov,kresicleftright,kresicorth} and for superalgebras in Ref. \onlinecite{kresic:super}.
In the present paper, we review Weyl realizations of Lie deformed spaces and corresponding star products, as well as the twist corresponding to Weyl realization and coproduct of momenta. 
We point out that in the original Snyder space~\cite{Snyder} there exists the symmetric ordering, but there is no Weyl realization and the star product is non associative.
For quadratic deformations of Minkowski space considered in Section 6, the symmetric ordering~(\ref{eq:sym})
is deformed and a generalized Weyl realization is defined.

In Section 2, Weyl realizations of Lie deformed spaces and corresponding star products are presented. In Section 3, the twist corresponding to
Weyl realization and coproduct of momenta is constructed. In Section 4, some examples for Weyl realizations of Lie deformed spaces are given.
In Section 5, it is shown that  for the original Snyder model there exists the symmetric ordering, but no Weyl realization exists. In Section 6,
quadratic deformations of the Minkowski space are considered and it is demonstrated that the symmetric ordering must be deformed and a generalized
Weyl realization can be defined. Finally, an Appendix related to Section 2 is presented.

\section{Weyl realization of Lie-deformed quantum Minkowski spaces and star product}

Let $\ggx$ be a finite dimensional Lie algebra over the field $\C$ with ordered basis $\hat x_0, \hat x_1,\hat x_2, \ldots, \hat x_{n-1}$
satisfying the commutation relations
\begin{equation}\label{1.01}
[\hat x_\mu,\hat x_\nu]= i l\, C_{\mu\nu\al} \hat x_\al
\end{equation}
where $i=\sqrt{-1}$ and $l C_{\mu\nu\al}$ are the structure constants of $\ggx$.
Since $[\hat x_\mu, \hat x_\nu]\to 0$ as $l\to 0$, the Lie algebra $\ggx$ is interpreted as a deformation of the underlying commutative space.
The enveloping algebra $U(\ggx)$ is then viewed as a noncommutative (NC) space of Lie algebra type generated.
Hence, it is of interest from both mathematical and physical point of view to study the NC coordinates $\hat x_\mu$ as deformations of the
commutative coordinates $x_\mu$. Let $\mathcal{H}_n$ denote the unital, associative algebra generated by
$x_\mu,p_\mu$, $0\leq \mu\leq n-1$, satisfying the commutation relations
\begin{equation}
[x_\mu,x_\nu]=[p_\mu,p_\nu]=0, \quad [p_\mu,x_\nu]=-i\eta_{\mu\nu}
\end{equation}
where $\eta=\text{diag}(-1,1,\ldots, 1)$ is the Minkowski metric. The subalgebra generated by $x_\mu$'s is the coordinate algebra of the
Minkowski space $M_n$. The Minkowski space $M_n$ itself by definition consists of points $q=(q_0,q_1,\ldots,q_{n-1})$ where all $q_i$ are real
numbers and which is equipped with Minkowski metric. The algebra $\mathcal{H}_n$ is an undeformed phase space with commutative coordinates
$x_\mu$ and momenta $p_\mu$. Throughout the paper, summation over repeated indices is defined with respect to the Minkowski metric,
$A_\al B_\al =\sum_{\al, \beta=0}^{n-1} A_\al\, \eta_{\al \beta}\, B_\beta$. A realization of NC coordinates $\hat x_\mu$ is given by
\begin{equation}\label{2-4}
\hat x_\mu = x_\al\, \varphi_{\al\mu}(l,p)
\end{equation}
where $\varphi_{\al\mu}(l,p)$ is a formal power series in $p_0,p_1,\ldots, p_{n-1}$ such that $\lim_{l\to 0} \varphi_{\al\mu}(l,p)=\eta_{\al\mu}$.
This guarantees that $\hat x_\mu \to x_\mu$ as $l\to 0$.
The commutation relations in $\ggx$ imply that the functions $\varphi_{\al\mu}(l,p)$ satisfy a system of coupled partial differential equations.
Such systems are usually under-determined and have infinitely many solutions depending on arbitrary analytic functions.

Denote by $\gg$ the underlying vector space of $\ggx$ and by $S(\gg)$ the symmetric algebra on $\gg$. While the basis $\hat x_0, \hat x_1 \ldots, \hat x_{n-1}$ of $\ggx$
is also a set of algebra generators of $U(\ggx)$, the corresponding generators in the symmetric algebra $S(\gg)$ are denoted
by $x_0, x_1, \ldots, x_{n-1}$ without $\hat{}$ symbol to emphasize that
they commute. Then $S(\gg)$ can be viewed as a commutative subalgebra of $\mathcal{H}_n$ consisting of polynomials in $x_0,x_1,\ldots,x_{n-1}$.
Define a left action $\rhd \colon \mathcal{H}_n \otimes S(\gg) \to S(\gg)$ of $\mathcal{H}_n$ on $S(\gg)$ by
\begin{equation} \label{2-5}
x_\mu \rhd f(x) = x_\mu f(x), \quad p_\mu \rhd f(x) = -i\frac{\p f}{\p x_\mu}, \quad 1\rhd f(x)=f(x).
\end{equation}
A realization \eqref{2-4}, i.e.\ the choice of the functions $\varphi_{\al\mu}$, is related to an ordering on the enveloping algebra of $\ggx$ as follows.
Define the linear map $\Omega_\varphi \colon U(\ggx)\to S(\gg)$ by $\Omega_\varphi (\hat f) =\hat f\rhd 1$ where $\hat f$ on the right-hand side is
computed in realization $\varphi$. As a consequence of the Poincar\'{e}--Birkhoff--Witt theorem, $\Omega_\varphi$ is an isomorphism of vector spaces.
The Weyl symmetric realization is the realization $\varphi$ such that $\Omega_\varphi^{-1} \colon S(\gg)\to U(\ggx)$ is the symmetrization map,
i.e.
\begin{equation}
\Omega_\varphi^{-1} (x_{i_1} x_{i_2} \ldots x_{i_k})=\frac{1}{k!} \sum_{\sigma\in\Sigma_k} \hat x_{i_{\sigma(1)}} \hat x_{i_{\sigma(2)}}\ldots \hat x_{i_{\sigma(k)}}.
\end{equation}
This property of $\varphi$ is equivalent with
\begin{align}\label{2-6}
(k\hat x)^N \rhd 1 &= (kx)^N \notag \\
&= \sum_{\begin{array}{c}m_0+\ldots+m_{n-1}=N
    \\ m_0, m_1, \ldots, m_{n-1}\geq 0\end{array}}
\binom{N}{m_0 m_1 \ldots m_{n-1}}
    (k_0 x_0)^{m_0}\cdots (k_{n-1} x_{n-1})^{m_{n-1}},
\end{align}
for all $N\geq 1$, where $k\hat x= k_\al\, \hat x_\al$ and $kx = k_\al\, x_\al$. The Weyl realization for a general Lie algebra was discussed in Refs. \onlinecite{petracci,Durov,kresicleftright}
where it was shown that it can be constructed as follows.
Let $\mathcal{C}(p)$ denote a matrix of order $n$ with elements $\mathcal{C}(p)_{\mu\nu}=C_{\al\mu\nu} p_\al$. Then
the Weyl realization is given by
\begin{equation}\label{eq:uf1}
\hat x_\mu = x_\al \left(\frac{l\mathcal{C}(p)}{1-e^{-l\mathcal{C}(p)}}\right)_{\mu\al} = x_\al\, \psi(l\mathcal{C}(p))_{\mu\al}
\end{equation}
where $\psi(t)=t/(1-e^{-t})$ is the generating function for the Bernoulli numbers and $\psi(l\mathcal{C}(p))$ is a matrix of order $n$ with elements
\begin{equation}\label{eq:uf2}
\psi(l\mathcal{C}(p))_{\mu\nu} = \left(\frac{l\mathcal{C}(p)}{1-e^{-l\mathcal{C}(p)}}\right)_{\mu\nu}.
\end{equation}
As a consequence of property \eqref{2-6}, in this realization
\begin{equation}
e^{ik \hat x}\rhd 1 = e^{ikx} \quad \text{and} \quad p_\mu e^{ik\hat x} \rhd 1 = k_\mu e^{ikx}.
\end{equation}
The Weyl realization for $\kappa$-deformed Euclidean space
was found in Ref. \onlinecite{meljkjst} and the
Weyl realization for the orthogonal, Lorentz and Poincar\'e algebras was investigated in Ref. \onlinecite{kresicorth}.
Generalization of the Weyl realization to a class
of Lie superalgebras was given in Ref. \onlinecite{kresic:super}.

Another important concept related to realizations of a Lie algebra $\ggx$ is the star product on the symmetric algebra $S(\gg)$. One can introduce an associative product on $S(\gg)$, denoted
$\ast$, such that the vector space isomorphism $\Omega_\varphi^{-1} \colon S(\gg)\to U(\ggx)$ becomes an isomorphism of associative algebras. The star product is defined by
\begin{equation}\label{eq:starprod}
f\ast g = \Omega_\varphi \big(\Omega_\varphi^{-1}(f) \Omega_\varphi^{-1}(g)\big), \quad f,g\in S(\gg).
\end{equation}
If we denote $\hat f = \Omega_\varphi^{-1}(f)$ and $\hat g = \Omega_\varphi^{-1}(g)$, then the star product can be written using the action \eqref{2-5} as
\begin{equation}\label{2-11}
f\ast g = (\hat f \hat g)\rhd 1 = \hat f \rhd g.
\end{equation}
Then from the definition of the star product we have $\Omega_\varphi (\hat f \hat g) = \Omega_\varphi (\hat f)\ast \Omega_\varphi (\hat g)$. Associativity of the
multiplication in $U(\ggx)$ implies that the star product is also associative. Clearly, the star product depends on the
realization $\varphi$ and the structure of the Lie algebra $\ggx$. However, for monomials of order one in any realization we have
\begin{equation}
x_\mu \ast x_\nu - x_\nu \ast x_\mu = i l C_{\mu\nu\al}\, x_\al.
\end{equation}
By using the formal completion of $S(\gg)$, the star product \eqref{eq:starprod} is readily extended to analytic functions in $x_0,x_1,\ldots, x_n$.
In the Weyl realization, the star product of exponential functions yields
\begin{equation}\label{eq:stsymn1}
e^{ikx}\ast e^{iqx} = \big(e^{ik \hat x} e^{iq\hat x}\big) \rhd 1 = e^{iD(k,q)x}
\end{equation}
where the function $\DD\colon M_n \times M_n\to M_n$ is given by
\begin{equation}\label{eq:stsymn2}
\DD_\mu(k,q) = \exp\left(k_\beta \, \psi(l\mathcal{C}(q))_{\beta\al} \,\frac{\p}{\p q_\al}\right) q_\mu
\end{equation}
and $\mathcal{C}(q)$ is a matrix with elements $\mathcal{C}(q)_{\mu\nu}= C_{\al\mu\nu}\, q_\al$. Solution for the function $\DD_\mu(k,q)$ is found using
Theorem 1 in Ref. \onlinecite{mstrajnexp21}, and it describes the Baker-Campbell-Hausdorff series in an alternative way.

Equations \eqref{eq:stsymn1} and \eqref{eq:stsymn2} define deformed addition of momenta by $(k\oplus q)_\mu= \DD_\mu(k,q)$.
For example, to the third order in $l$
we have (see Appendix)
\begin{align}\label{eq:thirdorderDkq}
\DD_\mu(k,q) &= k_\mu + q_\mu + \frac{l}{2} k_\al\, \mathcal{C}(q)_{\al\mu} \notag \\
&+ \frac{l^2}{12} \Big[ k_\al\, (\mathcal{C}(q)^2)_{\al\mu} + q_\al\, (\mathcal{C}(k)^{2})_{\al\mu}\Big] \notag \\
&- \frac{l^3}{48} \Big[k_\al\, (\mathcal{C}(q)^{2})_{\al\beta}\, \mathcal{C}(k)_{\beta\mu}-q_\al\, (\mathcal{C}(k)^{2})_{\al\beta}\, \mathcal{C}(q)_{\beta\mu}\Big].
\end{align}
From associativity of the star product, it follows immediately that $\DD$ satisfies the associativity rule
\begin{equation}
\DD(\DD(k_1,k_2),k_3)=\DD(k_1,\DD(k_2,k_3)).
\end{equation}
If the structure constants $C_{\mu\nu\lambda}$ are
replaced by $-C_{\mu\nu\lambda}$, we obtain the Lie algebra $\ggx^{\mathrm{op}}$ with Lie bracket $[\hat x_\mu,\hat x_\nu]^{op}=[\hat x_\nu,\hat x_\mu]$. Then the star product
defined by Eq. \eqref{eq:starprod} for the Lie algebra $\ggx^{\mathrm{op}}$ is related to the star product for $\ggx$ by
\begin{equation}
f\ast_{op}g = g\ast f
\end{equation}
and the corresponding function $\DD^{\mathrm{op}}\colon M_n\times M_n\to M_n$ defined by the relation
\begin{equation}
e^{ikx}\ast_{\mathrm{op}} e^{iqx} = e^{i\DD^{\mathrm{op}}(k,q)x}
\end{equation}
is given by $\DD^{\mathrm{op}}(k,q) = \DD(q,k)$.

Many other aspects of the symmetric ordering and Weyl realization of Lie algebra type NC spaces were exhibited as the equivalent properties
of deformation quantization of linear Poisson structures~\cite{berezin1967,gutt,karasev,Kathotia}. In particular, Gutt~\cite{gutt} has shown how to
extend the star product to all smooth functions on the cotangent bundle of a Lie group.
This star product corresponds
to the star product~(\ref{eq:stsymn1}),(\ref{eq:stsymn2}),(\ref{eq:thirdorderDkq}), where $\hat{x}$ is expressed
in terms of the universal formula~(\ref{eq:uf1})--(\ref{eq:uf2})
related to the symmetric ordering (as explicitly seen in Ref. \onlinecite{Durov}, Ch. 10).
Kathotia~\cite{Kathotia} has studied (in terms of a diagrammatic expansion) the difference between the star products in two quantizations of
the linear Poisson structures: the star product in symmetric ordering and the Kontsevich star product. Petracci~\cite{petracci} has worked out
the universal representation by coderivations (which corresponds to the Weyl realization) of finite dimensional Lie algebras and Lie superalgebras. In Ref. \onlinecite{snote},
a symmetry property shared by the Weyl realization of Lie algebras and
by a slightly more general class of realizations
of associative algebras has been proven.

\section{Twist corresponding to Weyl realization and coproduct of momenta}

In this section, we consider Drinfeld-Xu twist elements related to the Weyl (symmetric) realization of a Lie algebra $\g$ and the associated star product.
For a general Lie algebra, Heisenberg algebras and their deformations, representing some noncommutative phase spaces, fail to have a Hopf algebra
structure needed for the classical Drinfeld twist formalism, but do have a more general Hopf algebroid structure.
The structure of a Hopf algebroid~\cite{bohmHbk,xu} on
$\kappa$-deformed noncommutative phase spaces
and the corresponding Drinfeld-Xu twists~\cite{xu} were found in Refs. \onlinecite{jurmeljs,jurkovmelj}, see also Refs. \onlinecite{meljjord2017,lukmeljpikwor}.
The Hopf algebroid structure on the phase spaces
for general Lie algebra type noncommutative spaces is exhibited in Ref. \onlinecite{halg}
and the correspoding twists were found in Ref. \onlinecite{meljsk2018}.

Let us first recall the notion of a Drinfeld twist for Hopf algebras.
For a Hopf algebra $H$ with coproduct $\Delta$ and counit $\varepsilon$, a Drinfeld twist is an invertible element
$\F\in H\otimes H$ satisfying
\begin{align}
(\F\otimes 1) (\Delta \otimes\id) \F &= (1\otimes \F)(\id\otimes \Delta) \F \quad \text{(2-cocycle condition)}, \label{3-18} \\
(\varepsilon \otimes\id) \F &= (\id \otimes \varepsilon) \F = 1 \quad \text{(normalization condition)}.
\end{align}
Given a Hopf action $\rhd\colon H\otimes A\to A$ on an associative algebra $A$ (which is therefore a left $H$-module algebra)
and a twist $\F$ on $H$, we can form a new algebra structure on $A$ with multiplication
\begin{equation}\label{3-20}
a\ast b = m\circ \F^{-1} \rhd (a\otimes b) = \sum_\al (\bar f^\al \rhd a) (\bar f_\al \rhd b),
\end{equation}
where we symbolically write $\F=\sum_\al f^\al \otimes f_\al$ and
$\F^{-1} = \sum_\al \bar f^\al \otimes \bar f_\al$, and $m$
is the original multiplication on $A$. The 2-cocycle condition ensures that the star product \eqref{3-20} is associative and the normalization
condition implies that $1\ast a = a\ast 1 = a$. Moreover, the twist induces a twisted coproduct on $H$ which makes it again into a Hopf algebra with the same multiplication.

Natural examples give rise to a more general notion of a bialgebroid twist.
The notion of a bialgebroid/Hopf algebroid $\mathcal{H}$ over a possibly noncommutative base algebra $U$ generalizes the notion a
bialgebra/Hopf algebra over a field. A (left) $U$-bialgebroid~\cite{xu,bohmHbk,halg,meljsk2018} is an algebra $\mathcal{H}$ which is a
$U$-bimodule in a specific way and equipped with a map of $U$-bimodules, the counital coassociative coproduct
$\Delta:\mathcal{H}\to\mathcal{H}\otimes_U\mathcal{H}$ where $\otimes_U$ is the tensor product of $U$-bimodules.
Coassociativity of the coproduct $\Delta$ is understood as an identity
$$
(\Delta\otimes_U\id)\circ\Delta = (\id\otimes_U\Delta)\circ\Delta
$$
among the elements in the bimodule tensor product $\mathcal{H}\otimes_U\mathcal{H}\otimes_U\mathcal{H}$~\cite{meljsk2018} and it must admit a
counit which is a map of $U$-bimodules $\varepsilon\colon\mathcal{H}\to U$.
Additional compatibilities between these structures are required~\cite{xu,bohmHbk,halg}.
A Hopf algebroid is a bialgebroid equipped with an antipode map.
The Drinfeld-Xu twist for a $U$-bialgebroid $\mathcal{H}$ is an element
$\F\in\mathcal{H}\otimes_{U}\mathcal{H}$ satisfying the 2-cocycle condition (which is now an identity in $\mathcal{H}\otimes_U\mathcal{H}\otimes_U\mathcal{H}$),
normalization and invertibility, all adapted to the setting of tensor products of $U$-bimodules\footnote{The notation and the roles of $\F$ versus $\F^{-1}$
for Drinfeld-Xu twists are here interchanged with respect to Ref. \onlinecite{xu} and Ref. \onlinecite{meljsk2018}.}. In our example $\F \in \mathcal{H}_n\hat\otimes_{U(\ggx)} \mathcal{H}_n$
and the tensor product is appropriately completed.
A twist $\F$ for the Lie algebra type noncommutative phase space is given by
\begin{equation}\label{2-22}
\F=\exp\big(-ip_\beta \otimes \hat x_\beta\big) \exp\big(ip_\al \otimes x_\al\big),
\end{equation}
where $\hat x_\beta$ is the Weyl realization~\eqref{eq:uf1} of the Lie algebra $\ggx$. One can show that the generators of $\ggx$
can be written as
\begin{equation}
\hat x_\mu = m\big(\mathcal{F}^{-1} (\rhd \otimes\id) (x_\mu \otimes 1)\big),
\end{equation}
where $m$ denotes multiplication in $S(\gg)$, the action $\rhd$ is defined by Eq. \eqref{2-5} and
\begin{equation}\label{eq:Fm}
\mathcal{F}^{-1} = \exp(-ip_\al \otimes x_\al) \exp(ip_\beta \otimes \hat x_\beta).
\end{equation}
If we write formally $\F=\exp(f)$, then we can expand $f =\ln(\F)$ in a form
which is schematically, supressing the indices and numerical coefficients, given by
\begin{equation}
  f = \sum_{k,r=1}^\infty p^k\otimes C^{k+r-1} x p^r =\sum_{k=0}^\infty\sum_{r=0}^\infty
  p^k\otimes C^{k+r}L p^r
\end{equation}
where $L_{\mu\nu}=x_\mu p_\nu$ are the generators of $gl(n)$. For example, to second order in the structure constants of $\ggx$ the expansion truncates to
\begin{align}
f &= -i p_\al \otimes \Big(\frac{1}{2} x_\beta\, l\mathcal{C}(p)_{\al\beta}+\frac{1}{12} x_\beta \big(l\mathcal{C}(p)\big)^2_{\al\beta}\Big) + \frac{l^2}{2} p_\al\, p_\beta \otimes
\frac{i}{12} x_\gamma\, \mathcal{C}(p)_{\beta\delta}\, C_{\delta \al \gamma} \notag  \\
&= \frac{il}{2} \mathcal{C}(p)_{\gamma\beta} \otimes L_{\beta\gamma} + \frac{il^2}{12} \mathcal{C}(p)_{\gamma \delta} \otimes L_{\beta\gamma}\, \mathcal{C}(p)_{\beta\delta}
-\frac{il^2}{24} \big(\mathcal{C}(p)\big)^2_{\lambda \gamma} \otimes L_{\gamma\lambda}.
\end{align}

We note that $\mathcal{H}_n$ does not have the structure of a Hopf algebra.
The commutative subalgebra $\mathcal{P}\subset\mathcal{H}_n$
generated by the momenta $p_0, p_1, \ldots, p_{n-1}$
has a Hopf algebra structure where the momenta are primitive elements,
that is,
\begin{equation}
\Delta_0(p_\mu) = 1\otimes p_\mu + p_\mu \otimes 1, \quad S_0(p_\mu)=-p_\mu, \quad \varepsilon_0 (p_\mu)=0.
\end{equation}
The twist $\F$ defined by \eqref{2-22} is not a Drinfeld twist since it does not satisfy the 2-cocycle condition \eqref{3-18}. However, $\F$ does
satisfy the 2-cocycle condition in the bialgebroid sense~\cite{xu}, as shown in Ref. \onlinecite{meljsk2018}.
The Hopf algebra structure of $\mathcal{P}$ can be
twisted by $\F$ by retaining the old multiplication and twisting
the coproduct to a deformed coassociative coproduct $\Delta$ given by
\begin{equation}\label{3-28}
\Delta (P) = \F \Delta_0 (P) \F^{-1},
\end{equation}
for any $P\in\mathcal{P}$. Essentially the same formula, appropriately understood (the twist changes the multiplication of the noncommutative
base algebra $U$ over which the bimodule structure on $\mathcal{H}$ and the bimodule tensor product $\otimes_U$ are defined)
can be used more generally for any $P$ in $\mathcal{H}$.
The deformed coproduct
expresses the Leibniz rule for the action of $P$ in $\mathcal{P}$ on
the star product in the corresponding realization, $m_\ast(\Delta(P)(\triangleright\otimes\triangleright)(f\otimes g)) = P\triangleright(f\ast g)$.
To third order in the structure constants of $\ggx$, the deformed coproduct of momenta is
\begin{align}
\Delta (p_\mu) &= p_\mu \otimes 1 + 1\otimes p_\mu + \frac{l}{2} p_\al \otimes \mathcal{C}(p)_{\al\mu} \notag \\
&+ \frac{l^2}{12}\big(p_\al \otimes (\mathcal{C}(p))^2_{\al\mu}+(\mathcal{C}(p))^2_{\al\mu}\otimes p_\al\big) \notag \\
&+\frac{l^3}{48}\big(\mathcal{C}(p)_{\al\beta}\otimes p_\al\,
\mathcal{C}(p)_{\beta\mu}-p_\al \, \mathcal{C}(p)_{\beta\mu} (\mathcal{C}(p))^2_{\al\beta}\big).
\end{align}
The antipode is given by the undeformed formula, $S(p_\mu)=-p_\mu$.
Furthermore, using the deformed coproduct \eqref{3-28} one finds that the Weyl realization of NC coordinates $\hat{x}_\mu$ can be written as
\begin{equation}
\hat x_\mu = x_\mu + ix_\al m\circ (\Delta-\Delta_0)(p_\al) (\Delta\otimes\id)(x_\mu \otimes 1).
\end{equation}
The twist element \eqref{eq:Fm} can also be expressed in terms of the deformed coproduct $\Delta$ which yields the identities~\cite{meljjord2017}
\begin{equation}\label{3-30}
  \F^{-1}=\colon \exp\big(i(1\otimes x_\al) (\Delta-\Delta_0)(p_\al)\big)\colon
  = \exp(-ip_\alpha\otimes x_\alpha)\exp(ip_\beta\otimes\hat{x}_\beta)
\end{equation}
and $\Delta p_\mu= \DD_\mu(p\otimes 1,1\otimes p)$,
where $\colon \, \colon$ denotes the normal ordering
with $x$'s to the left and $p$'s to the right.

\subsection{Opposite Lie algebra with $C\mapsto -C$}

Interchanging the order of elements in the Lie bracket amounts to changing the sign of structure constants. Thus, we consider a Lie algebra
$U(\ggy)$ with generators $\hat{y}_0,\ldots,\hat{y}_{n-1}$
satisfying
\begin{equation}
  [\hat y_\mu, \hat y_\nu] = -i l C_{\mu\nu\al} \hat y_\al.
\end{equation}
The antiisomorphism $\hat x_\alpha\mapsto \hat y_\alpha$ from $\ggx$ to $\ggy$ makes every left module of $U(\ggx)$ into a right module of $U(\ggy)$
and maps the tensor product of  $U(\ggx)$-bimodules by interchanging the order of tensor factors (for the representatives) to the tensor product of $U(\ggx)$-bimodules in the opposite order.

In particular, by interchaging the tensor factors in $\F =  \exp(-i p_\beta \otimes \hat x_\beta)\exp(ip_\al \otimes x_\al)$ we obtain
\begin{equation}
\F^{\mathrm{op}} = \exp(-i \hat x_\beta \otimes p_\beta)\exp(ix_\al \otimes p_\al).
\end{equation}
where the tensor product is over $U(\ggy)$.

Let us define the realization of $\hat{y}_\mu$ within the same Heisenberg algebra by
\begin{equation}\label{3.31}
\hat y_\mu = m\big((\F^{\mathrm{op}})^{-1} (\rhd \otimes\id) (x_\mu \otimes 1)\big).
\end{equation}
Within this realization,
\begin{equation}
[\hat x_\mu, \hat y_\nu]=0.
\end{equation}
These realizations can be interpreted geometrically (see Ref. \onlinecite{halg}, Sections 2 and 3) where the Lie algebra is realized by left invariant and,
alternatively, by right invariant vector fields on a Lie group which obviously commute.

Furthermore, in the Weyl realization of $U(\ggy)$, the formula for
the generators $\hat{y}_\mu$,
\begin{equation}
\hat y_\mu = x_\alpha\left(\frac{l\mathcal{C}(p)}{e^{l\mathcal{C}(p)}-1}\right)_{\mu\al}
\end{equation}
is obtained from \eqref{eq:uf1}  by changing the structure constants $C$ to $-C$. Therefore,
\begin{equation}
\hat x_\mu = m \big((\F^{\mathrm{op}}_{-C})^{-1} (\rhd \otimes\id) (x_\mu \otimes 1)\big)
\end{equation}
where
\begin{equation}\label{eq:Fcm}
(\F^{\mathrm{op}}_{-C})^{-1} := \exp(-i x_\al \otimes p_\al) \exp(i\hat y_\beta \otimes p_\beta).
\end{equation}
The original coproduct of momenta \eqref{3-28} is given in terms of
the opposite twist as
\begin{equation}
\Delta (p_\mu) = \F^{\mathrm{op}}_{-C}\, \Delta_0 (p_\mu)\, (\F^{\mathrm{op}}_{-C})^{-1} \equiv \Delta^{\mathrm{op}}_{-C} (p_\mu).
\end{equation}
The twists $\F_{-C}^{\mathrm{op}}$ and $\F$ are given by different formulas, but they generate the same realization of $\hat x_\mu$.
While the right-hand sides of the formula~(\ref{eq:Fm}) for $\F^{-1}$ and of the formula (\ref{eq:Fcm}) for $(\F^{\mathrm{op}}_{-C})^{-1}$
understood as elements in the tensor product $\mathcal{H}_n\otimes\mathcal{H}_n$ over the field are different, the twists $\F^{-1}$
and $(\F^{\mathrm{op}}_{-C})^{-1}$ should be understood as the projection of these expressions into the bimodule tensor product
$\mathcal{H}_n\otimes_{U(\ggx)}\mathcal{H}_n$ where they coincide~\cite{meljsk2018}. While $\F$ is a twist for a left bialgebroid,
$(\F^{\mathrm{op}}_{-C})^{-1}$ can be understood as a twist for a right bialgebroid; taking the opposite base algebra, which is
equivalent to changing the sign in the structure constants, interchanges these the two twists.
The same observations hold for the twists corresponding to (different) realizations of the Lie algebra $gl(n)$ exhibited in Ref. \onlinecite{gln}.

\section{Examples of Weyl realizations}

\subsection{Kappa-Minkowski space}

The $\kappa$-Minkowski space is a Lie algebra type NC space defined by commutation relations
\begin{equation}\label{4-38}
[\hat x_\mu, \hat x_\nu] = i(a_\mu \hat x_\nu - a_\nu \hat x_\mu), \quad 0\leq \mu,\nu \leq n-1,
\end{equation}
where $a\in M_n$. The structure constants of the Lie algebra \eqref{4-38} are given by $C_{\mu\nu\lambda}=a_\mu \eta_{\nu\lambda}-a_\nu \eta_{\mu\lambda}$.
The $\kappa$-Minkowski space
provides a framework for doubly special relativity theory~\cite{kowgliknow} and it has applications in quantum gravity and quantum field theory. Using Eq.~\eqref{eq:uf1}
we find
\begin{equation}
\hat x_\mu = x_\mu \varphi_S(A) - a_\mu (xp) \frac{1-\varphi_S(A)}{A}
\end{equation}
where $A=-a_\al p_\al$, $xp=x_\al p_\al$ and
\begin{equation}
\varphi_S(A) = \frac{A}{e^A-1}.
\end{equation}
The commutation relations between the NC coordinates and momenta are found to be
\begin{equation}
[p_\mu, \hat x_\nu] = -i\eta_{\mu\nu} \varphi_S(A)+ia_\nu p_\mu \frac{1-\varphi_S(A)}{A}
\end{equation}
and the deformed coproduct of momenta is given by
\begin{equation}
\Delta (p_\mu) = \varphi_S(\Delta_0 (A)) \left(\frac{p_\mu}{\varphi_S(A)}\otimes 1\right)+\varphi_S(-\Delta_0 (A)) \left(1\otimes \frac{p_\mu}{\varphi_S(-A)}\right)
\end{equation}
where $\Delta_0 = A\otimes 1 + 1\otimes A$. The identity \eqref{3-30}, where $\F^{-1}$ is expressed by Eq.~\eqref{eq:Fm}, holds for the above realization of $\hat x_\mu$ and
coproduct $\Delta (p_\mu)$. The Weyl realizations of Lie algebra type NC spaces isomorphic to the orthogonal algebra $so(n)$, Lorentz algebra $so(1,n-1)$
and Poincar\'{e} algebra can be found in Ref. \onlinecite{kresicorth}.

\subsection{NC spaces with extended tensorial coordinates}

Let us consider the Minkowski space extended with tensorial coordinates $\hat x_{(\mu\nu)}$ satisfying the commutation relations~\cite{everton,amorim}
\begin{equation}\label{4-43}
[\hat x_\mu,\hat x_\nu]=il\, \hat x_{(\mu\nu)}, \quad [\hat x_\mu, \hat x_{(\al\beta)}]=0, \quad [\hat x_{(\mu\nu)},\hat x_{(\al\beta)}]=0,
\end{equation}
and $\hat{x}_{(\mu\nu)}=-\hat{x}_{(\nu\mu)}$. Here
$l$ plays the role of a minimal length, and $\hat x_{(\mu\nu)}$ transform as a second rank antisymmetric tensor under the Lorentz
generators $M_{\al\beta}$.
The structure constants of the Lie algebra \eqref{4-43} are given by
\begin{alignat}{2}
C_{\mu\nu\lambda} &=0, & \qquad C_{\mu\nu (\al\beta)} &= \frac{1}{2}(\eta_{\al\mu}\, \eta_{\beta\nu}-\eta_{\al\nu}\, \eta_{\beta\mu}), \\
C_{\mu (\al\beta) \lambda} &=0, & \qquad C_{\mu (\al\beta) (\gamma \delta)} &=0, \\
C_{(\al\beta) (\gamma\delta) \mu} &=0, & \qquad C_{(\al\beta)(\gamma\delta)(\rho\sigma)} &=0.
\end{alignat}
The Weyl realization of the algebra \eqref{4-43} is
\begin{equation}
  \hat x_\mu =  x_\mu -\frac{l}{2}x_{\mu\al}p_\al, \quad \hat x_{\mu\nu} = x_{\mu\nu},
\end{equation}
where the algebra generated by $x_\mu$, $p_\mu$, $x_{\mu\nu}$ and $p_{\mu\nu}$ is defined by commutation relations
\begin{equation}
  [p_\mu,x_\nu]=-i\eta_{\mu\nu}, \quad
  [x_{\mu\nu},p_{\rho\sigma}] =
  i(\eta_{\mu\rho}\eta_{\nu\sigma} - \eta_{\mu\sigma}\eta_{\nu\rho})
\end{equation}
and the remaining commutators vanish,
$$
  [x_{(\mu\nu)},p_\lambda]=0, \quad
  [x_{\mu},p_{\rho\sigma}] =0,\quad
  [p_{(\mu\nu)},p_{(\rho\sigma)}]=0.
$$
  Furthermore, relations
  \begin{equation}
  [p_\mu,\hat{x}_\nu] =-i\eta_{\mu\nu},\quad
  [p_{\mu\nu},\hat x_\lambda]=\frac{il}{2}(\eta_{\mu\lambda}p_\nu-\eta_{\nu\lambda}p_\nu)
  \end{equation}
  and $e^{ik\hat{x}}\triangleright 1 = e^{ikx}$ hold.
  Coproducts of the momenta are given by
  \begin{align}
    \Delta p_\mu &= \Delta_0 p_\mu = p_\mu\otimes 1+1\otimes p_\mu \\
    \Delta p_{\mu\nu} &= p_{\mu\nu}\otimes 1+1\otimes p_{\mu\nu}+\frac{il}{2}(p_\mu\otimes p_\nu-p_\nu\otimes p_\mu)
  \end{align}
  The realization of $\hat x_\mu$ can also be written using the twist element
  \begin{equation}
\F^{-1}=\exp\left(-\frac{il}{2}x_{(\alpha\beta)}p_\alpha\otimes p_\beta\right)
  \end{equation}
which yields
\begin{equation}
\hat x_\mu = m(\F^{-1} (\rhd \otimes\id)(x_\mu \otimes 1)) = x_\mu - \frac{l}{2} x_{\mu\al}p_\al.
\end{equation}

Additional tensorial coordinates in extended Snyder model were proposed in Ref. \onlinecite{girliv} and the Heisenberg double for extended Snyder model was constructed in Ref. \onlinecite{pachdouble}.
Weyl realizations of extended Snyder space and $\kappa$--deformed extended Snyder space can be found in Refs. \onlinecite{meljmig1,meljmig2,meljmig3}.

\subsection{Canonical $\theta_{\mu\nu}$-deformation}

Consider the algebra~(\ref{eq:thetagen}) in the simple case when $\theta_{\mu\nu}$ are constants antisymmetric in $\mu,\nu$,
\begin{equation}
  [\hat x_\mu,\hat x_\nu ] = i\theta_{\mu\nu},
  \quad
  [p_\mu,\hat x_\nu ] = -i\eta_{\mu\nu},
  \quad [p_\mu, p_\nu ] = 0.
\end{equation}
The NC coordinates $\hat{x}_\mu$  do not close a Lie algebra, but the Weyl realization and symmetric ordering can be defined similarly as in Section 4.2.
The Weyl realization is
\begin{equation}
  \hat{x}_\mu = x_\mu - \frac{1}{2}\theta_{\mu\alpha}p_\alpha
\end{equation}
and $\exp(ik\hat{x}) \triangleright 1 = e^{ikx}$ holds.
Furthermore, the twist is given by 
\begin{equation}
\mathcal{F}^{-1}=\exp(\frac{-i}{2}\theta_{\alpha\beta}p_\alpha\otimes p_\beta).
\end{equation}

\section{Snyder space with symmetric ordering and without Weyl realization}

The Snyder space~\cite{Snyder} is defined by commutation relations
\begin{align}
[\hat x_\mu, \hat x_\nu] &= i\beta M_{\mu\nu},  \label{5-50} \\
[M_{\mu\nu},\hat x_\lambda] &= i(\eta_{\mu\lambda} \hat x_\nu - \eta_{\nu \lambda} \hat x_\mu), \\
[M_{\mu\nu},M_{\rho\sigma}] &= i(\eta_{\mu\rho} M_{\nu\sigma} - \eta_{\mu\sigma} M_{\nu\rho}-\eta_{\nu\rho} M_{\mu\sigma} + \eta_{\nu\sigma} M_{\mu\rho}).  \label{5-52}
\end{align}
The NC coordinates $\hat x_\mu$ and Lorentz generators $M_{\mu\nu}$ close a Lie algebra $\gg_{\hat{x},M}$. However, the generators $M_{\mu\nu}$ cannot be interpreted as NC
coordinates since NC coordinates must satisfy the condition $\hat x_\mu \rhd 1 = x_\mu$ while
\begin{equation}
M_{\mu\nu} \triangleright 1 = (x_\mu p_\nu - x_\nu p_\mu)\triangleright 1= 0.
\end{equation}
Therefore, the formula for the Weyl realization \eqref{eq:uf1} cannot be applied to $\hat{x}_\mu$ and $M_{\mu\nu}$. We can define the
enveloping algebra $U(\gg_{\hat x,M})$ of the Lie algebra \eqref{5-50}--\eqref{5-52} and the projection $U(\gg_{\hat x,M})\rhd 1 = U(\gg_x)$ with
nonassociative star product~\cite{battisti1,girliv}. The coproduct of momenta is noncoassociative and the corresponding twist does not satisfy the cocycle condition.

Realizations of $\hat x_\mu$ were found in Refs. \onlinecite{battisti1,battisti2},
\begin{equation}\label{5-54}
\hat x_\mu = x_\mu \varphi_1 (\beta p^2) + \beta\, (xp)\, p_\mu\, \varphi_2(\beta p^2),
\end{equation}
where $\varphi_1$ is an arbitrary function satisfying $\varphi_1(0)=1$ and
\begin{equation}\label{eq:varphi2}
\varphi_2 = \frac{1+2 \varphi_1^\prime \varphi_1}{\varphi_1 - 2 \beta \, p^2\, \varphi_1^\prime}
\end{equation}
where $\varphi_1^\prime$ denotes the derivative of $\varphi_1$ with respect to $\beta\, p^2$. If $\hat x_\mu$ is given by the realization \eqref{5-54}, then
\begin{equation}
e^{ik_\alpha\hat x_\alpha} \rhd 1 = e^{iK_\alpha (k) x_\alpha}
\end{equation}
where the functions $K_\al(k)$ satisfy a system of differential equations~\cite{mstrajnexp21}
\begin{eqnarray}
 k_\beta\, \frac{\p K_\mu (k)}{\p k_\beta} = k_\beta\, \varphi_{\mu\beta}\big(K(k)\big),
 \\
 \label{eq:dif2}
\frac{\p K_\mu (\lambda k)}{\p \lambda} = k_\beta\, \varphi_{\mu\beta}\big(K(\lambda k)\big).
\end{eqnarray}
The corresponding coproduct of momenta is obtained from
\begin{equation}
\Delta (p_\mu) = \DD_\mu (p\otimes 1, 1\otimes p) = \exp\left(K^{-1}_\beta (p)\otimes \varphi_{\al\beta} (p) \frac{\p}{\p p_\al}\right) (1\otimes p_\mu)
\end{equation}
where $K^{-1}(k) = (K^{-1}_\mu (k))_\mu$ is the inverse vector function of
$K(k) = (K_\mu(k))_\mu$,
\begin{equation}
K^{-1}_\mu (K(k)) = K_\mu (K^{-1}(k)) = k_\mu.
\end{equation}
The twist element is given by
\begin{align}
\F^{-1} &= \,\,\colon\!\exp\left((1\otimes x_\al) (\Delta - \Delta_0)\, p_\al\right) \colon \\
&= \exp\left(-i(p_\al\otimes x_\al)\right) \exp\left(K_\beta^{-1} (p)\otimes \hat x_\beta\right).
\end{align}
Note that there exists a realization of $\hat{x}_\mu$ corresponding to the
symmetric ordering $e^{ik\hat{x}}\triangleright 1 = e^{ikx}$, i.e.\ with $K_\mu(k) = k_\mu$.
Substituting this into~(\ref{eq:dif2}) we get
\begin{equation}
\varphi_2(\beta p^2) = \frac{1-\varphi_1(\beta p^2)}{\beta p^2}.
\end{equation}
Taking into account~(\ref{eq:varphi2}) and defining $u = \beta p^2$, we obtain the
differential equation
\begin{equation}
  2 u\varphi_1'(u) =
  \varphi_1(u)-(\varphi_1(u))^2-u
\end{equation}
whose solution is
\begin{equation}
\varphi_1(u) = \sqrt{u}\, \cot(\sqrt{u}).
\end{equation}
Therefore, the realization of $\hat{x}_\mu$ corresponding to the symmetric ordering, $K_\mu(k) = k_\mu$ is
\begin{equation}
\hat x_\mu = x_\mu + (x_\mu p^2 - (xp) p_\mu ) \frac{\varphi_1 (\beta\, p^2)-1}{p^2} = x_\mu\left(1-\frac{\beta}{3}p^2\right) +\frac{\beta}{3}(xp)p_\mu + O(\beta^2).
\end{equation}

\section{Quadratic deformations of Minkowski space with generalized Weyl realization}

In Ref. \onlinecite{Wess} $q$-deformed Heisenberg algebras were considered as examples of noncommutative structures. A framework for higher dimensional
NC spaces based on quantum groups was studied. Furthermore, quadratic deformations of Minkowski space from twisted Poincar\'{e} symmetries were
considered in Ref. \onlinecite{lukwor}.

A quadratic algebra is usually defined~\cite{polishchuk,manin} as the tensor algebra $T(V)$ on a finite dimensional vector space $V$ modulo the
ideal generated by some subspace of the second tensor power $T^2(V)$ of $V$. We limit our attention to a class of quadratic algebras described as
follows. A quadratic algebra $U_\theta$ is the tensor algebra on $V = \operatorname{Span}_{\mathbb{C}}\{\hat x_0,\hat x_1\ldots,\hat{x}_{n-1}\}$
modulo the 2-sided ideal $I_\theta\subset T(V)$
generated by the commutation relations
\begin{equation}\label{eq:thetaquad}
[\hat x_\mu, \hat x_\nu] = \theta_{\mu\nu\gamma\delta} \hat x_\gamma \hat x_\delta,
\end{equation}
where $\theta_{\mu\nu\gamma\delta}$ are real constants antisymmetric in $\mu,\nu$,
and satisfy a nondegeneracy condition
that the linear span
of the set of relations~(\ref{eq:thetaquad})
is of maximal dimension,
\begin{equation}
  \dim_{\mathbb{C}}\operatorname{Span}_{\mathbb{C}}
  \{
  \hat{x}_\mu\hat{x}_\nu-\hat{x}_\nu\hat{x}_\mu-
  \theta_{\mu\nu\gamma\delta}\hat x_\gamma \hat x_\delta
  \,|\, \mu < \nu \} = n(n-1)/2,
\end{equation}
just like in the commutative case when
$\theta_{\mu\nu\gamma\delta}=0$. We also require that all Jacobi identities for the commutators $[\hat x_\mu, \hat x_\nu]$
replaced by the right-hand side in~(\ref{eq:thetaquad})
hold already in the tensor algebra\footnote{Jacobi identity is trivially satisfied for the usual bracket $[a,b]=ab-ba$ in any associative algebra.
However, if we view the right-hand side in~(\ref{eq:thetaquad}),
  $\theta_{\mu\nu\gamma\delta}\hat{x}_\gamma\hat{x}_\delta\in T^2(V)$,
as a formula for a new bilinear bracket denoted $\{\hat{x}_\mu,\hat{x}_\nu\}$,
then one can impose a {\it nontrivial} Jacobi identity on this bracket $\{,\}$.
The bracket $\{\{\hat{x}_\mu, \hat{x}_\nu\},\hat{x}_\tau\} =
\theta_{\mu\nu\gamma\delta}\{\hat{x}_\gamma\hat{x}_\delta,\hat{x}_\tau\}$
involves third order expressions
$\{\hat{x}_\gamma\hat{x}_\delta,\hat{x}_\tau\}$ in $T^3(V)$
defined using the Leibniz rule
$\{ab,c\}= \{a,c\}b+a\{b,c\}$, hence obtaining
$\{\{\hat{x}_\mu, \hat{x}_\nu\},\hat{x}_\tau\}
= (\theta_{\mu\nu\gamma\delta}\theta_{\gamma\tau\rho\sigma}+
\theta_{\mu\nu\rho\gamma}\theta_{\gamma\tau\sigma\delta})
\hat{x}_\rho\hat{x}_\sigma\hat{x}_\delta$. We should warn the reader that
the bracket $\{,\}$ cannot be extended to the entire tensor algebra $T(V)$
as a Lie bracket satisfying the Leibniz rule (nonabelian Poisson bracket).
Namely, the only such bracket on
$T(V)$ is the usual commutator (see Ref. \onlinecite{farkas}, Theorem 1.2).}.
Then the Jacobi identity for any triple of generators
$\hat{x}_\mu,\hat{x}_\nu,\hat{x}_\tau$ reduces to a
condition that for any $\rho,\sigma,\delta$ we have
\begin{equation}\label{eq:Jacobithetaquad}
  \theta_{\mu\nu\gamma\delta}\theta_{\gamma\tau\rho\sigma}+
  \theta_{\mu\nu\rho\gamma}\theta_{\gamma\tau\sigma\delta}+
  \theta_{\nu\tau\gamma\delta}\theta_{\gamma\mu\rho\sigma}+
  \theta_{\nu\tau\rho\gamma}\theta_{\gamma\mu\sigma\delta}+
  \theta_{\tau\mu\gamma\delta}\theta_{\gamma\nu\rho\sigma}+
  \theta_{\tau\mu\rho\gamma}\theta_{\gamma\nu\sigma\delta} = 0.
\end{equation}
Identities~(\ref{eq:Jacobithetaquad}) play an important role in the
study of quadratic local Poisson and double Poisson structures,
see Ref. \onlinecite{rubtsov}. We view the associative algebra $U_\theta=T(V)/I_\theta$ as an analogue
of the universal enveloping algebra of a Lie algebra.
Rescaling
the structure constants  $\theta_{\mu\nu\rho\sigma}$ by a common formal
parameter we obtain a family of algebras $U_\theta$.
When the parameter is set to $0$
we obtain the symmetric algebra $S(V)$. We would like that this
family of algebras is in fact a deformation of the commutative algebra $S(V)$.
That means that an isomorphism of vector spaces $U_\theta\cong S(V)$ is fixed
for all values of the parameter
and this isomorphism equals to the identity map when the parameter is $0$.
To support this deformation picture we propose the following construction.
As we did with the enveloping algebras $U(\ggx)$
for the Lie algebra type deformations before,
we first embed the algebra $U_\theta$ via a realization map
 $\hat{f}\mapsto \hat{f}_\varphi$
into the (possibly completed) Heisenberg algebra $\mathcal{H}_n$.
Then we project the result into $S(V)$ by the Fock action~(\ref{eq:Fock}),
$\hat{f}_\varphi\mapsto \hat{f}_\varphi\triangleright 1$,
on the unit element in $S(V)$.
The composition of the two steps,
$\hat{f}\mapsto \hat{f}_\varphi\triangleright 1$,
is a linear map $\Omega_\varphi\colon U_\theta\to S(V)$
by construction. This map may or may not be invertible,
depending on the quadratic algebra and its realization $\varphi$.
If $\Omega_\varphi$ is indeed an isomorphism of vector spaces,
then using formula~(\ref{eq:starprod}),
we can transfer the product from $U_{\theta}$ to $S(V)$
along $\Omega_\varphi$ to obtain an associative star product $\ast$ on $S(V)$.
In this way we exhibit an algebra with the underlying space $S(V)$ which is
isomorphic as an algebra to $U_\theta$. Note that
the generators of the undeformed algebra $S(V)$ are denoted by $x_\alpha$
without $\hat{}$ symbol which is compatible with the notation for
the undeformed Heisenberg algebra $\mathcal{H}_n$.

A general class of realizations of NC coordinates $\hat x_\mu\in U_\theta$
is of the form
\begin{equation}
\hat x_\mu = x_\al\, \varphi_{\al\mu}(iL_{\gamma\delta})
\end{equation}
where $L_{\al\beta} = x_\al\, p_\beta$ generate the Lie algebra $gl(n)$,
\begin{equation}
[L_{\mu\nu},L_{\rho\sigma}] = i(\eta_{\mu\sigma} L_{\rho\nu} - \eta_{\rho\nu} L_{\mu\sigma}).
\end{equation}

A generalized Weyl realization can be constructed in a similar manner as in Ref. \onlinecite{Durov} starting with the following expression in the lowest order in $\theta_{\mu\nu\gamma\delta}$,
\begin{equation}
\hat x_\mu = x_\mu + \frac{i}{2} \theta_{\mu\gamma\beta\al} \, x_\al x_\beta\, p_\gamma + O(\theta^2).
\end{equation}
We point out that the results for Lie algebra type NC spaces discussed in Sections 2 and 3 cannot be applied to these quadratic deformations.
Construction of such quadratic algebras can be performed using twist operators that besides the Lorentz generators $M_{\mu\nu}$ include dilation operators
and, more generally, $L_{\mu\nu}=x_\mu p_\nu$. One example of such twist is (see Ref. \onlinecite{ms2112})
\begin{equation}
\F = \exp\Big(\sum_{\al,\beta} a_{\al\beta} D_\al \otimes D_\beta\Big), \quad a_{\beta\al} = - a_{\al\beta}
\end{equation}
where $D_\al = x_\al p_\al$ (no summation). Another example is given in Ref. \onlinecite{lukwor},
\begin{equation}
\F = \exp\Big(\frac{i}{2} \theta_{\al\beta\gamma\delta} M_{\al\beta} \wedge M_{\gamma\delta}\Big)
\end{equation}
where
\begin{equation}
\theta_{\al\beta\gamma\delta} = - \theta_{\beta\al\gamma\delta} = - \theta_{\al\beta\delta\gamma}=-\theta_{\gamma\delta\al\beta}
\end{equation}
for any different, but fixed indices $\alpha,\beta,\gamma,\delta$.
If, for a quadratic algebra, we define functions $K(k,x)$
and $D(k,q,x)$ by the identities
\begin{align}
e^{ik\hat x}\rhd 1 &= e^{iK(k,x)}, \\
e^{ikx} \ast e^{iqx} &= e^{iD(k,q,x)},
\end{align}
then $K(q,x)$ and $D(k,q,x)$ are generally not linear in $x$.
If the twist is Abelian, then the star product is associative and the coproduct of momenta is coassociative. The coproduct of momenta
$\Delta (p_\mu) = \F \Delta_0 (p_\mu) \F^{-1}$
depends, besides on momenta, on the Lorentz generators $M_{\mu\nu}$, dilations $D_\mu$ and, more generally, on the generators $L_{\al\beta}$ as well.
Furthermore,
\begin{equation}
(f\ast g) (x) = m\big(\F^{-1} (\rhd \otimes \rhd)(f(x)\otimes g(x))\big)
\end{equation}
and
\begin{equation}
\hat x_\mu = m\big(\F^{-1} (\rhd \otimes\id)(x_\mu \otimes 1)\big).
\end{equation}

If the twist is given by
\begin{equation}
\F = \exp\big(\sum_{\al,\beta} a_{\al,\beta}D_\al \otimes D_\beta\big),
\end{equation}
we find
\begin{equation}\label{6-77}
\hat x_\al = x_\al \exp\big(\sum_\beta i a_{\al\beta} D_\beta\big)
\end{equation}
which yields
\begin{equation}
\hat x_\al \hat x_\beta = q_{\al\beta}^2\, \hat x_\beta \hat x_\al
\end{equation}
and
\begin{equation}
x_\al \ast x_\beta = q_{\al\beta}^2\, x_\beta \ast x_\al,
\end{equation}
where $q_{\al\beta} := e^{a_{\al\beta}}$. In this example we use the convention $\alpha,\beta\in\{1,\ldots,n\}$.
We note that $q_{\al\beta}\, q_{\beta\al} = 1$ and $q_{\al\al}=1$ (no summation). In this case, the generalized
symmetric ordering is defined as
\begin{equation}
  \big(k_\al \hat x_\al\big)^N \rhd 1 = \sum_{\begin{array}{c}m_1, m_2, \ldots m_n\\
  m_1+\ldots+m_n=N\end{array}}\binom{N}{m_1,m_2,\ldots m_n}_q (k_1 x_1)^{m_1} \ldots (k_n x_n)^{m_n}
\end{equation}
and
\begin{equation}
e^{ik \hat x}\rhd 1 = (e^{ikx})_q
\end{equation}
where
\begin{equation}\label{6-82}
\binom{N}{m_1, m_2, \ldots m_n}_q
\end{equation}
are $q_{\al\beta}$--deformed multinomial coefficients and $(e^{ikx})_q$ denotes the $q_{\al\beta}$--deformed exponential function~\cite{koornSpecial}. For example,
\begin{equation}
\hat x_{\al_1} \hat x_{\al_2} \ldots \hat x_{\al_N} \rhd 1 = \Pi_{\substack{\al_k, \al_l \\ k<l}}\, q_{\al_k \al_l}\, x_1^{m_1} x_2^{m_2} \ldots x_n^{m_n}
\end{equation}
with $\sum_{i=1}^n m_i = N$ we get $q_{\al\beta}$--deformed multinomial coefficient \eqref{6-82}.
In particular, for $N=2$
\begin{equation}
(\hat x_1 \hat x_2 + \hat x_2 \hat x_1)\rhd 1 = (q_{12}+q_{21}) x_1 x_2.
\end{equation}
For $N=3$ we have
\begin{align}
&(\hat x_1 \hat x_2 \hat x_3 + \hat x_2 \hat x_3 \hat x_1+\hat x_3 \hat x_1 \hat x_2+\hat x_2 \hat x_1 \hat x_3+\hat x_1 \hat x_3 \hat x_2+\hat x_3 \hat x_2 \hat x_1)\rhd 1 \\
&=(q_{12} q_{13} q_{23}+q_{23} q_{21} q_{31}+q_{31} q_{32} q_{12} + q_{21} q_{23} q_{13}+q_{13} q_{12}q_{32}+q_{32} q_{31} q_{21}) x_1 x_2 x_3
\end{align}
and similarly
\begin{align}
(\hat x_1 \hat x_1 \hat x_2 + \hat x_1 \hat x_2 \hat x_1 + \hat x_2 \hat x_1 \hat x_1) \rhd 1 &= (q_{11} q_{12} q_{12} + q_{12} q_{21} q_{11}+q_{21} q_{21} q_{11}) x_1^2 x_2 \\
&=\big(q_{12}^2 + 1 + q_{12}^{-2}) x_1^2 x_2  \label{6-88}
\end{align}
If $a_{\al\beta}=a$, $\al<\beta$, then $q_{\al\beta} = q$ and $q_{\beta\al}=q^{-1}$,
and the above results
reduce to $q$-commutation relations in Refs. \onlinecite{koornSpecial,koornqFour}.

\section{Appendix}

In this appendix we track explicitly
the dependence of
structure constants and realizations on a small parameter $l$,
hence we shall use the convention where
$[\hat{x}_\alpha,\hat{x}_\beta]= il C_{\alpha\beta\lambda}\hat{x}_\lambda$.
The Weyl realization for a Lie algebra type noncommutative space is then~\cite{Durov}
\begin{equation}\tag{A1}
\hat{x}_\mu = x_\alpha\left( \frac{l \CC}{1 - e^{-l \CC}}\right)_{\mu\alpha} =
x_\mu + \frac{l}{2}x_\alpha \CC_{\mu\alpha}(p)+\frac{l^2}{12}x_\alpha(\CC^2)_{\mu\alpha}
+ O(l^4), 
\end{equation}
where we introduce the matrix $\CC = (C_{\mu\nu})$
by $\CC_{\mu\nu} = \CC(p)_{\mu\nu} = C_{\alpha\mu\nu}p_\alpha$.
Then there exists a unique function of two variables $D(k,q)$
(see Ref. \onlinecite{mstrajnexp21}, Theorem 2 (iii)) such that
\begin{equation}\tag{A2}
 e^{ik\cdot\hat{x}}e^{i q\cdot\hat{x}} = e^{i D(k,q)\cdot\hat{x}},  
\end{equation}
\begin{equation}\tag{A3}
\frac{d\DD_\mu(t k,q)}{dt} = k_\alpha\left(\frac{l\CC (s)}{1 - e^{-l\CC (s)}}\right)_{\alpha\mu}  
\end{equation}
where $s=D(tk,q)$ and $\DD_\mu$ satisfies the condition $\DD_\mu(0,q)=q_\mu$. In the limit $l = 0$, we obtain $\DD_\mu^{(0)}(k,q) =k_\mu+q_\mu$.
We solve the equation perturbatively order by order in $l$.
In the zeroth order in $l$, $\DD_\mu^{(0)}(tk,q)=tk_\mu+q_\mu$.
In the first order in $l$,
\begin{equation}\tag{A4}
  \frac{d D^{(1)}(tk,q)}{dt} =
  k_\mu + \frac{l}{2}k_\alpha C_{\beta\alpha\mu}\DD_\beta^{(0)}(tk,q) =
  k_\mu + \frac{l}{2}k_\alpha C_{\beta\alpha\mu}(tk_\beta+q_\beta).  
\end{equation} 
From here we find
\begin{equation}\tag{A5}
  D^{(1)}_\mu(tk,q) = tk_\mu + q_\mu + \frac{l}{2}tk_\alpha\CC(q)_{\alpha\mu}.  
\end{equation}
In the second order in $l$,
\begin{equation}\tag{A6}
  \frac{d D^{(2)}_\mu(tk,q)}{dt} =
  k_\mu + \frac{l}{2}k_\alpha C_{\beta\alpha\mu}D^{(1)}_\beta(tk,q)
  + \frac{l^2}{12}k_\alpha\CC^2_{\alpha\mu},  
\end{equation}
with the solution
\begin{equation}\tag{A7}
  D^{(2)}_\mu(tk,q) = tk_\mu+q_\mu + \frac{l}{2}tk_\alpha\CC(q)_{\alpha\mu}
  + \frac{l^2}{12}\left( tk_\alpha(\CC(q)^2)_{\alpha\mu}
  + t^2(\CC(k)^2)_{\alpha\mu}q_{\alpha} \right).
\end{equation}
In the third order in $l$,
\begin{equation}\tag{A8}
  \frac{d D^{(3)}(tk,q)}{dt} =k_\mu + \frac{l}{2}k_\alpha C_{\beta\alpha\mu}D^{(2)}_\beta(tk,q)
  + \frac{l^2}{12}k_\alpha\CC^2_{\alpha\mu},
\end{equation}
hence
\begin{equation}\begin{array}{ccc}\tag{A9}
    D^{(3)}_\mu(tk,q) &=& tk_\mu+q_\mu + \frac{l}{2}tk_\alpha\CC(q)_{\alpha\mu}+
    \frac{l^2}{12}\left(tk_\alpha(\CC(q)^2)_{\alpha\mu}+
    t^2(\CC(k)^2)_{\alpha\mu} q_\alpha\right)+ \\
   && + \frac{l^3}{48}t^2\left((\CC(k)^2)_{\alpha\beta}q_\alpha\CC(q)_{\beta\mu}
   - k_\alpha\CC(k)_{\beta\mu}(\CC(q)^2)_{\alpha\beta}\right).
  \end{array}
\end{equation}
In the Weyl realization, $\DD_\mu(q,k)$ does not change
if we replace $\CC$ by $-\CC$ and also the following formula holds
\begin{equation}\label{eq:D2part}\tag{A10}
  \DD_\mu(tk,q) = q_\mu +tk_\alpha\left(\frac{l\CC(q)}{1-e^{-l\CC(q)}}\right)_{\alpha\mu}
+\sum_{k=1}^\infty\frac{t^{k+1}}{(k+1)!}\left.\left(\frac{d^k}{dt^k}\psi_\mu(t)\right)\right|_{t=0}
\end{equation}
where $\psi$ is defined by
\begin{equation}\tag{A11}
\psi_\mu(t) = \frac{d \DD_\mu(tk,q)}{dt}.
\end{equation}
Notice that the right-hand side of~(\ref{eq:D2part})
is written a sum of the part not depending on $k$,
the part linear in $k$ and the remainder involving $\psi_\mu$.
This precisely corresponds to writing the Baker-Campbell-Hausdorff series
$H(X,Y) = \mathrm{ln}(e^X e^Y)$ as a sum of three summands: $Y$,
the part linear in $X$, and the higher order remainder. The linear part
is related to the differential of the exponential map at the unit of the
corresponding Lie group~(see Ref. \onlinecite{halg}, Section 3).
The simplified structure of~(\ref{eq:D2part}) for $t=1$ can be written as
\begin{equation}\tag{A12}
\DD_\mu(k,q) = q_\mu + k_\alpha\left(\frac{l\CC(q)}{1-e^{-l\CC(q)}}\right)_{\alpha\mu}+O(k^2).
\end{equation}

\section*{Acknowledgements}
We thank Rina \v{S}trajn for checking the formula~(\ref{eq:thirdorderDkq}).
Z.\v{S}. thanks V.~Rubtsov and G. Sharygin for discussions on quadratic algebras.
Z. \v{S} has been partly supported by the Croatian Science Foundation under the Project ``New Geometries for Gravity and Space'' (IP-2018-01-7615). 
\section*{Data Availability}
Data sharing is not applicable to this article as no new data were created or analyzed in this study.

\end{document}